\documentclass[structabstract]{aa}
\usepackage{graphicx}
\usepackage{txfonts}
\usepackage{longtable}
\usepackage{lscape}
\usepackage{natbib}
\begin{document}
   \title{First T dwarfs in the VISTA Hemisphere Survey \thanks{Based on observations 
made with the Calar Alto 3.5-m telescope, the Magellan telescope at Las Campanas,
the ESO Very Large Telescope at the Paranal Observatory, and the IAC80 at Teide
Observatory}}
   \subtitle{}

   \author{N. Lodieu \inst{1,2}
          \and
   B.\ Burningham \inst{3}
   \and
   A.\ Day-Jones \inst{4}
   \and
   R.--D.\ Scholz \inst{5}
   \and 
   F.\ Marocco \inst{3}
   \and
   S.\ Koposov \inst{6,7}
   \and 
   D.\ Barrado y Navascu\'es \inst{8,9}
   \and 
   P.\ W.\ Lucas \inst{3}
   \and 
   P.\ Cruz \inst{9}
   \and
   J.\ Lillo \inst{9}
   \and
   H.\ Jones \inst{3}
   \and
   A.\ Perez-Garrido \inst{10}
   \and 
   M.\ T.\ Ruiz \inst{4}
   \and 
   D.\ Pinfield \inst{3}
   \and 
   R.\ Rebolo \inst{1,2}
   \and 
   V.\ J.\ S.\ B\'ejar \inst{1,2}
   \and
   S.\ Boudreault \inst{1,2}
   \and
   J.\ P.\ Emerson \inst{11}
   \and
   M.\ Banerji \inst{9}
   \and
   E.\ Gonz\'alez-Solares \inst{6}
   \and
   S.\ T.\ Hodgkin \inst{6}
   \and
   R.\ McMahon \inst{6}
   \and
   J.\ Canty \inst{3}
   \and
   C.\ Contreras \inst{3}
          }

   \offprints{N. Lodieu}

   \institute{Instituto de Astrof\'isica de Canarias (IAC), Calle V\'ia L\'actea s/n, E-38200 La Laguna, Tenerife, Spain\\
         \email{nlodieu@iac.es}
         \and
         Departamento de Astrof\'isica, Universidad de La Laguna (ULL),
E-38205 La Laguna, Tenerife, Spain
         \and
         Centre for Astrophysics Research, Science and Technology Research Institute, University of Hertfordshire, Hatfield AL10 9AB, United Kingdom
         \and
         Departamento de Astronomia, Universidad de Chile, Casilla 36-D, Santiago, Chile
         \and
         Leibniz-Institut f\"ur Astrophysik Potsdam (AIP), An der Sternwarte 16, 14482, Potsdam, Germany
         \and
         Institute of Astronomy, Madingley Road, Cambridge CB3 0HA, United Kingdom
         \and
         Sternberg Astronomical Institute, Lomonosov Moscow State University, Universitetsky pr.\ 13, Moscow, 119992 Russia 
         \and
         Centro Astron\'omico Hispano Alem\'an, Calle Jes\'us Durb\'an Rem\'on 2-2, 04004 Almer\'ia, Spain
         \and
         Departamento de Astrofisica, Centro de Astrobiologia (INTA-CSIC), ESAC Campus, PO Box 78, E-28691 Villanueva de la Ca\~nada, Spain
         \and
         Universidad Polit\'ecnica de Cartagena, Campus Muralla del Mar, Cartagena, Murcia E-30202, Spain
         \and
         School of Physics \& Astronomy, Queen Mary University of London, Mile End Road, London E1 4NS, United Kingdom                  
}

   \date{\today{}; \today{}}

% \abstract{}{}{}{}{} 
% 5 {} token are mandatory
 
  \abstract
   {}
   {The aim of the project is to improve our current knowledge of the density
of T dwarfs and the shape of the substellar initial mass function by 
identifying a magnitude-limited sample of T dwarfs in the full southern sky.}
   {We present the results of a photometric search aimed at discovering cool 
brown dwarfs in the Southern sky imaged at infrared wavelengths by the
Visible and Infrared Survey Telescope for Astronomy (VISTA) and the Wide 
Infrared Survey Explorer (WISE) satellite mission. We combined the first 
data release (DR1) of the VISTA Hemisphere Survey (VHS) and the WISE
preliminary data release to extract candidates with red mid-infrared colours
and near- to mid-infrared colours characteristics of cool brown dwarfs.}
   {The VHS DR1 vs.\ WISE search returned tens of T dwarf candidates, 13 of 
which are presented here, including two previously published in the literature
and five new ones confirmed spectroscopically with spectral types between 
T4.5 and T8\@. We estimate that the two T6 dwarfs lie within 16 pc and the 
T4.5 within 25 pc. The remaining three are 30--50 pc distant. The only 
T7 dwarf in our sample is the faintest of its spectral class with 
$J$\,=\,19.28 mag. The other six T dwarf candidates remain without 
spectroscopic follow-up. We also improve our knowledge on the proper motion 
accuracy for three bright T dwarfs by combining multi-epoch data from 
public databases (DENIS, 2MASS, VHS, WISE, Spitzer).}
   {}


   \titlerunning{First T dwarfs in VHS}
   \maketitle
%
%________________________________________________________________

%
%%%%%%%%%%%%%%%%%%%%%%%%%%
%%%%% Introduction %%%%%
%%%%%%%%%%%%%%%%%%%%%%%%%%
%
\section{Introduction}
\label{VHS_dT:intro}

Brown dwarfs represent an important link between the coolest stars that burn
hydrogen and exoplanets that orbit stars. Our knowledge of their photometric 
and spectroscopic properties will help us improve current state-of-the-art
models and provide key information on the chemistry at play in cool atmospheres 
\citep{leggett07a,leggett09,burningham09,burgasser10c,burningham10a,leggett12a}.
Furthermore, brown dwarfs are important testing current star formation models 
via studying of the present-day mass function 
\citep{burgasser04d,metchev08,bate09,bastian10,burningham10b,reyle10,bate12,kirkpatrick12}.

The T spectral class consists of brown dwarfs with cool effective temperatures
\citep[1400--500\,K;][]{burgasser06a,burningham08a,lucas10,cushing11}.
The first T dwarf, Gl\,229B, was announced as a companion
to an M dwarf in 1995 by \citet{nakajima95}. Since then, more than 300 T dwarfs 
are known in the solar vicinity with a variety of physical properties 
\citep{kirkpatrick05}. The largest bulk of T dwarfs has been discovered in 
large-scale optical and infrared surveys such as the Two Micron All sky Survey
\citep[2MASS; e.g.][]{burgasser99}, the Sloan digital Sky Survey 
\citep[SDSS; e.g.][]{leggett00}, the UKIRT Infrared Deep Sky Survey
\citep[UKIDSS;][]{lawrence07} Large Area Survey 
\citep{kendall07,lodieu07b,pinfield08,chiu08,burningham10a,scholz10b,scholz12a},
the UKIDSS Galactic Clusters Survey \citep{lodieu09d}, the UKIDSS Deep 
Extragalactic Survey \citep{lodieu09b}, the CFHT Brown Dwarf Survey 
\citep{delorme08a,delorme08b,reyle10,albert11}, the Wide-F ield Infrared 
Survey Explorer \citep[WISE;][]{mainzer11,kirkpatrick11,scholz11,wright12}, 
and more recently Pan-Starrs \citep{deacon11,liu11b,deacon12a,deacon12b}.
The frontier between T and Y dwarfs originally proposed by 
\citet{kirkpatrick99} has now been crossed with the recent announcement of 
13 Y dwarfs by the WISE team \citep{cushing11,kirkpatrick12}.

The Visible and Infrared Survey Telescope for Astronomy 
\citep[VISTA;][]{emerson01,emerson04}, located at  the European Southern 
Observatory (ESO)'s Cerro Paranal Observatory in Chile, is a 4-m class 
telescope dedicated to imaging the southern sky. The telescope is equipped 
with the world's largest infrared camera, VIRCAM \citep{dalton06}, which is
composed of 67 million pixels offering an field-of-view 1.65 degrees in 
diameter, 0.6 square degrees of which are sampled by 'pawprints' of 16 
non-contiguous detectors with 0.34 arcsec pixels. Six suitably offset and 
jittered pawprints are 
combined into a filled 1.5 square degree 'tile', in which each piece of sky has 
been sampled by at least two pixels (ignoring the jitters), together with two 
additional 5.5$\times$88.5 arcmin strips at opposite sides of the tile that
are just covered once. Five broad-band filters are available, 
similar to the $ZYJHK$ installed on UKIRT/WFCAM \citep{hewett06} except for 
the $K_{s}$ filter. Two narrow-band filters are also available. 
Seventy-five percent 
of the time available on VISTA is dedicated to public surveys, one of which
is the VISTA Hemisphere Survey (VHS; PI McMahon, Cambridge, UK), with
which it is planned to 
image the entire southern sky in (at least) $J$ and $K_{s}$ over five years.
The main scientific aims of VHS include (1) examining the nearest and 
coolest stars and brown dwarfs, (2) studying the merger history of the Galaxy, 
(3) measuring the properties of dark energy through examining
large-scale structure to a redshift of $\sim$1, and (4) discovery of 
high-redshift quasars.

The ultimate scientific aim of our consortium is to identify a magnitude-limited
sample of T dwarfs to perform an initial mass function population analysis 
with robust 
statistics based on discoveries in the entire Southern Sky imaged within the 
framework of the VHS\@. Additional scientific interests include identifying 
rare T dwarfs within the large sample to study the effects of unusual 
properties (e.g.\ gravity and/or metallicity) on the observed spectra, 
conducting kinematic studies of populations over different temperature
ranges to constrain formation models, and revealing substellar halo objects.
In this paper we report the discovery of the first T dwarfs 
in the area (675 square degrees) common to the VHS DR1 (McMahon et al.\ 
2012, in prep) and the WISE preliminary data release \citep{wright10}.  
In Sect.\ \ref{VHS_dT:select} we describe the photometric 
selection criteria designed to identify cool T dwarfs. 
In Sects.\ \ref{VHS_dT:phot_obs} and \ref{VHS_dT:spec_obs} we detail 
the photometric and spectroscopic follow-up conducted at optical- and 
near-infrared wavelengths with various telescopes and instruments available 
to our team.
In Sect.\ \ref{VHS_dT:new_dT} we present the analysis of the spectra and 
derive the main spectral types and spectroscopic distances of the newly 
identified T dwarfs. Finally, we conclude and present our future plans
in Sect.\ \ref{VHS_dT:conclusions}.

\section{Sample selection}
\label{VHS_dT:select}

We cross-correlated the WISE preliminary data release with the VHS DR1 using a 
matching radius of 6 arcsec to take into consideration the resolution of the 
WISE images and include potential high proper motion sources despite the short
baseline ($<$\,1 year) between VHS and WISE\@. 
The common area includes 675 square degrees imaged in $JK_{s}$ by VHS, 
including 109 square degrees with Sloan coverage. Among those 675 square 
degrees, 276 and 186 are covered in $JHK_{s}$ and $YJHK_{s}$, respectively. 
The photometric and colour criteria detailed below were gathered into a 
structure query language query \citep{hambly08} launched in the VISTA Science 
Archive (Cross et al.\ 2012)\footnote{The VISTA Science Archive is at 
http://horus.roe.ac.uk/vsa/}. We selected only good-quality point sources 
outside the Galactic Plane (i.e.\ galactic latitudes higher than 15 degrees 
in absolute values) with at least ten WISE measurements (parameters 
{\tt{w1m,w2m,w3m,w4m}})\footnote{WISE provides at least eight frames over 99\% 
of the sky and a minimum of ten frames over 90\% of the sky \citep{wright10}} 
and high signal-to-noise detections in $w2$ and $J$ to focus specifically
on mid- to late-T dwarfs:

\begin{itemize}
\item Signal-to-noise WISE detections in $w2$: {\tt{w2snr}}$\geq$7
\item Error on the $J$ and $K_{s}$ photometry less than 0.16 and 0.3 mag, 
respectively (if $K_{s}$ magnitudes are quoted in the VHS catalogue)
\item ($w1-w2$) colours redder than 1.4 mag to focus on T dwarfs 
\citep{mainzer11,kirkpatrick11,kirkpatrick12}
\item ($J-w2$) colours redder than 1.9 mag \citep{kirkpatrick11,kirkpatrick12}
although we may miss some T4--T6 dwarfs due to the dip present in the
spectral type vs $J-w2$ relation shown in figure 7 of \citet{kirkpatrick11}
\item $J-K_{s}$\,$\leq$\,0.3 mag \citep{burgasser06a} or non-detection in $K_{s}$ 
\end{itemize}

The query returned 148 sources. The inspection of the WISE cut-outs allowed 
us to reduce the number of candidates to 64 sources after removing
artefacts and rejecting sources clearly detected in $w3$ and $w4$. 
Among the remaining 64 candidates, 18 were detected in SDSS DR8 within 5 
arcsec: 17 harbour blue $i-z$ colours inconsistent with cool T dwarfs (most 
appear as blue or dark blue sources on the SDSS colour images, suggesting that 
they might be white dwarfs, but none have spectra) and one shows a faint 
$z$-band detection \citep[$z$\,=\,20.87$\pm$0.21; AB system;][]{fukugita96} 
and no flux in the other Sloan bands (VHS J023756.24$-$063142.9).

\begin{figure*}
  \includegraphics[width=\linewidth, angle=0]{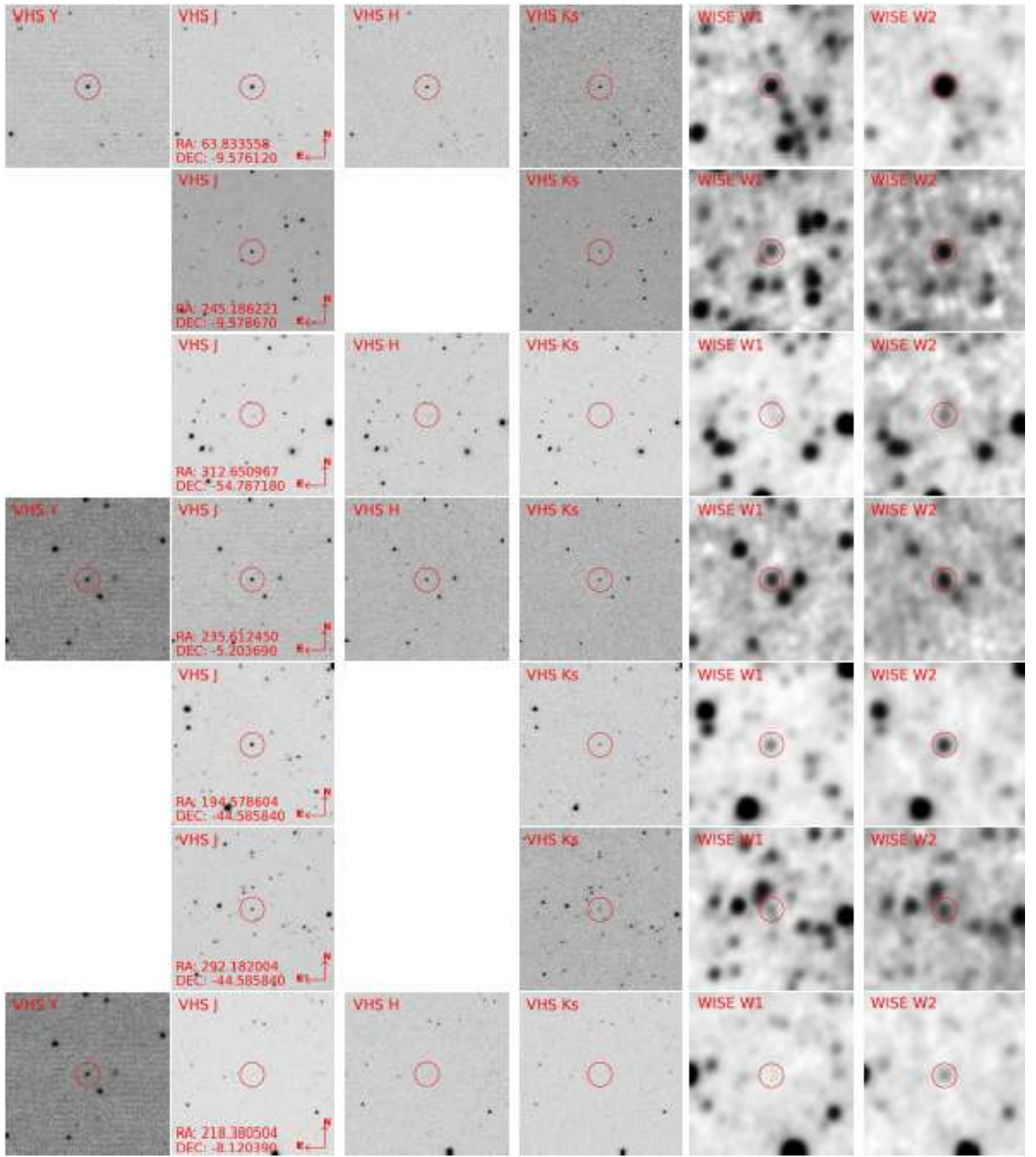}
  \caption{
Finding charts for the T dwarfs confirmed spectroscopically: we show
the VHS $YJHK_{s}$ and WISE $w1,w2$ images. Images are 2 arcmin a side with
north up and east to the left. The coordinates of the object are marked on 
the $J$-band chart.
}
  \label{fig_VHS_dT:fc_dT_1}
\end{figure*}

\begin{figure*}
  \includegraphics[width=\linewidth, angle=0]{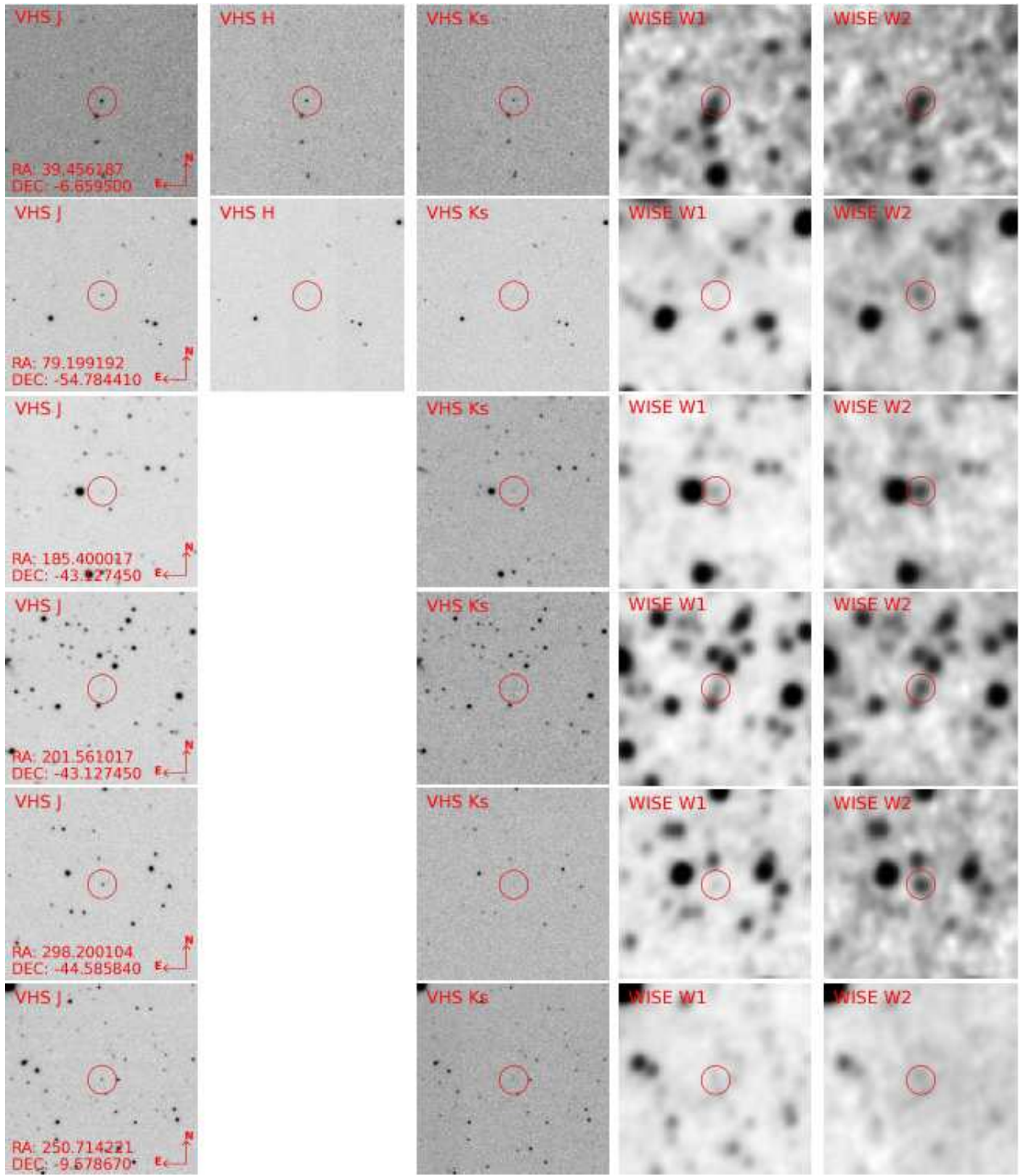}
  \caption{
Finding charts for the T dwarf candidates with out spectroscopic follow-up:
we show the VHS $JHK_{s}$ and WISE $w1,w2$ images (no $Y$ images are available
for these sources). Images are 2 arcmin a side with north up and east to the 
left. The coordinates of the object are marked on the $J$-band chart.
}
  \label{fig_VHS_dT:fc_dT_2}
\end{figure*}

Twenty-two sources (64$-$17) have USNO counterparts within 2 arcsec, reducing 
the final number of candidates to 25, including the 7 presented in this paper 
with spectroscopic confirmation and spectral types. None of the objects compiled
in the USNO catalogue has significant proper motion nor spectral type quoted 
in the {\tt{Simbad}} database, therefore they are either distant stars or
extragalactic sources.
Among the other candidates (25$-$7\,=\,18), four have $J-H$ colours redder
than 0.1 mag (including two with $J-K_{s}$\,$\geq$\,0.4 mag),
two are detected only in $K_{s}$ (likely M dwarfs), one has $Y-J$\,$\leq$\,0.5 
mag, one has $J-K_{s}$\,$\geq$\,0.4 mag, two show no VHS detection at the 
WISE position although a magnitude is quoted in the catalogue, 
and two have VHS images affected by bad pixels, making their candidacy dubious. 
The remaining six candidates (including VHS J023756.24$-$063142.9) remain 
interesting but we were unable to obtain spectra for them due to observing 
priorities and visibility issues.
The coordinates (J2000) of the seven confirmed T dwarfs and the six
candidates lacking spectroscopy are listed in Table \ref{tab_VHS_dT:tab_new_dT}
along with the near-infrared magnitudes from VHS DR1 and mid-infrared 
photometry from the WISE Preliminary Release which has now been superseded by 
the WISE All-sky release. Table \ref{tab_VHS_dT:tab_new_dT} lists spectral 
types accurate to half a subclass, spectroscopic distances, and instruments 
(FIRE, XSH stands for X-shooter) used to obtain the spectrum for our confirmed 
T dwarfs. We emphasise that 2MASSI J0415195$-$093506 was already known and 
has a trigonometric parallax suggesting a distance of of 5.736$\pm$0.092 pc 
\citep{burgasser02,vrba04} whereas VHS J162208.94$-$095934.6 was 
discovered independently by \citet{kirkpatrick11}. The last three columns of
Table \ref{tab_VHS_dT:tab_new_dT} give the proper motions
as well as the number of data points from UKIDSS, WISE, 2MASS, and DENIS used
to derive those values for the three brightest T dwarfs in our sample.
The finding charts of the objects presented in this paper are displayed in
Figures \ref{fig_VHS_dT:fc_dT_1} and \ref{fig_VHS_dT:fc_dT_2}.

\longtab{1}{\scriptsize
\begin{landscape}
 \begin{longtable}{@{\hspace{0mm}}c @{\hspace{1mm}}c @{\hspace{2mm}}c @{\hspace{2mm}}c @{\hspace{2mm}}c @{\hspace{2mm}}c @{\hspace{2mm}}c @{\hspace{2mm}}c @{\hspace{2mm}}c @{\hspace{2mm}}c @{\hspace{2mm}}c @{\hspace{2mm}}c @{\hspace{2mm}}c @{\hspace{2mm}}c @{\hspace{2mm}}c @{\hspace{2mm}}c@{\hspace{0mm}}}
 \caption{\label{tab_VHS_dT:tab_new_dT} Photometry in the Vega system for 
our sample of T dwarfs from VHS DR1 and the WISE Preliminary data release: 
objects with and without spectra are listed in the top and bottom panel.
} \\
 \hline
 \hline
RA (J2000)  & dec (J2000) & $I$ & $z$ & $Y$ & $J$ & $H$ & $K_{s}$ & $w1$ & $w2$ & SpT & dist & Inst.\ &  $\mu_{\alpha}cos{\delta}$ & $\mu_{\delta}$ &  $N_{\rm ep}$ \cr
 \hline
hh:mm:ss.ss & $^{\circ}$:$'$:$''$ & mag & mag & mag & mag & mag & mag     & mag  & mag &    & pc &  & mas/yr & mas/yr & \cr
 \hline
  \endfirsthead
  \caption{continued.} \\
  \hline
  \hline
RA (J2000)  & dec (J2000) & $I$ & $z$ & $Y$ & $J$ & $H$ & $K_{s}$ & $w1$ & $w2$ & SpT & dist & Inst.\ &  $\mu_{\alpha}cos{\delta}$ & $\mu_{\delta}$ &  $N_{\rm ep}$ \cr
 \hline
hh:mm:ss.ss & $^{\circ}$:$'$:$''$ & mag & mag & mag & mag & mag & mag     & mag  & mag &    & pc &  & mas/yr & mas/yr & \cr
  \hline
  \endhead
  \hline
  \endfoot
 04:15:21.21 &  $-$09:35:00.6 &   --- $\pm$ ---  &    --- $\pm$ ---  & 16.438$\pm$0.009 & 15.343$\pm$0.004 & 15.666$\pm$0.012 & 15.658$\pm$0.023 & 15.095$\pm$0.042 & 12.240$\pm$0.026 &  T8.0 &  5.83$\pm$1.26 & --- & ---  & ---  &  --- \cr
 12:58:04.89 &  $-$44:12:32.4 &   --- $\pm$ ---  &    --- $\pm$ ---  &   --- $\pm$ ---  & 15.999$\pm$0.011 &   --- $\pm$ ---  & 16.164$\pm$0.045 & 15.739$\pm$0.071 & 14.002$\pm$0.049 &  T6.0 & 14.45$\pm$1.80 & FIRE & $+$113.1$\pm$06.1 &  $-$178.4$\pm$09.1  &   11 \cr
 14:33:11.46 &  $-$08:37:36.3 &   --- $\pm$ ---  &    --- $\pm$ ---  & 20.141$\pm$0.252 & 19.139$\pm$0.154 & 19.217$\pm$0.142$^{d}$ &   --- $\pm$ ---  & 18.229$\pm$ ---  & 14.971$\pm$0.114 &  T8.0 & 33.48$\pm$10.17 & FIRE & ---  & ---  &  --- \cr
 15:43:52.78 &  $-$04:39:09.6 & 22.02$\pm$0.15$^{a}$ & 20.95$\pm$0.27$^{c}$ & 17.625$\pm$0.019 & 16.445$\pm$0.009 & 16.453$\pm$0.017 & 16.434$\pm$0.040 & 15.916$\pm$0.071 & 14.351$\pm$0.063 &  T4.5 &  22.09$\pm$1.21 & FIRE & $-$65.8$\pm$07.3  &  $-$14.9$\pm$05.0  &  17 \cr
 16:22:08.94 &  $-$09:59:34.6 & 20.86$\pm$0.50$^{b}$ & 20.72$\pm$0.11$^{c}$ &  --- $\pm$ ---  & 16.218$\pm$0.011 &   --- $\pm$ ---  & 16.448$\pm$0.155 & 16.251$\pm$0.102 & 14.016$\pm$0.054 &  T6.0 &  15.99$\pm$1.98 & XSH$^{b}$ & $+$76.9$\pm$11.5  &  $+$59.5$\pm$09.5  &  11 \cr
 19:29:34.18 &  $-$44:25:50.5 &   --- $\pm$ ---  &    --- $\pm$ ---  &   --- $\pm$ ---  & 17.129$\pm$0.022 &   --- $\pm$ ---  & 16.841$\pm$0.076 & 16.376$\pm$0.092 & 14.958$\pm$0.083 &  T4.0 & 31.58$\pm$1.72 & XSH & ---  & ---  &  --- \cr
 20:51:59.38 &  $-$55:08:43.9 &   --- $\pm$ ---  &    --- $\pm$ ---  &   --- $\pm$ ---  & 19.280$\pm$0.083 & 19.704$\pm$0.281 & 19.216$\pm$0.380 & 17.791$\pm$ ---  & 15.524$\pm$0.150 &  T7.0 &  51.10$\pm$11.42 & FIRE & ---  & ---  &  --- \cr
 \hline
 \hline
 02:37:56.24 &  $-$06:31:42.9 &   --- $\pm$ ---  &    --- $\pm$ ---  &   --- $\pm$ ---  & 17.555$\pm$0.019 & 17.588$\pm$0.046 & 17.549$\pm$0.080 & 17.080$\pm$0.170 & 15.481$\pm$0.152 &  ---  &  ---  & --- & ---  & ---  &  --- \cr
 05:18:59.64 & $-$54:49: 3.2 &   --- $\pm$ ---  &    --- $\pm$ ---  &  --- $\pm$ ---  & 18.554$\pm$0.057 & 19.116$\pm$0.220 &  --- $\pm$ ---  & 18.853$\pm$0.427 & 15.739$\pm$0.090 &  ---  &  ---  & --- & ---  & ---  &  --- \cr
 12:21:48.51 & $-$43:32:14.8 &   --- $\pm$ ---  &    --- $\pm$ ---  &  --- $\pm$ ---  & 18.837$\pm$0.144 &  --- $\pm$ ---  &  --- $\pm$ ---  & 16.321$\pm$0.092 & 14.784$\pm$0.075 &  ---  &  ---  & --- & ---  & ---  &  --- \cr
 13:25:47.18 & $-$43:40:41.9 &   --- $\pm$ ---  &    --- $\pm$ ---  &  --- $\pm$ ---  & 18.793$\pm$0.140 &  --- $\pm$ ---  &  --- $\pm$ ---  & 16.841$\pm$0.170 & 15.265$\pm$0.136 &  ---  &  ---  & --- & ---  & ---  &  --- \cr
 16:42:35.32 & $-$09:30:20.6 &   --- $\pm$ ---  &    --- $\pm$ ---  &  --- $\pm$ ---  & 18.677$\pm$0.076 &  --- $\pm$ ---  & 18.531$\pm$0.303 & 16.706$\pm$0.163 & 15.155$\pm$0.126 &  ---  &  ---  & ---$^{e}$ & ---  & ---  &  --- \cr
 19:55:50.12 & $-$44:15:14.2 &   --- $\pm$ ---  &    --- $\pm$ ---  &  --- $\pm$ ---  & 17.405$\pm$0.042 &  --- $\pm$ ---  &  --- $\pm$ ---  & 17.314$\pm$0.193 & 15.043$\pm$0.096 &  ---  &  ---  & --- & ---  & ---  &  --- \cr
 \hline
\end{longtable}
% \footnote{Notes on individual objects}
$^{a}$ 2$\sigma$ confidence detection limit in Johnson $I$ (Vega system) from CAMELOT \\
$^{b}$ Marginal detection in Johnson $I$ (Vega system) from CAMELOT; the Sloan
$i$ detection is 21.48$\pm$0.12 mag (AB system) \\
$^{c}$ 2$\sigma$ confidence detection limit in SDSS $z$ (AB system) from CAMELOT \\
%$^{d}$ 2$\sigma$ confidence detection limit in SDSS $z$ (AB system) from CAMELOT \newline
$^{d}$ Calar Alto 3.5-m/Omega2000 $H$-band photometry (Vega system) \\
$^{e}$ A low signal-to-noise spectrum (30 minutes on-source integration) 
obtained with the Gemini-North Near-Infrared Integral Field Spectrometer
(NIFS) indicates that it has a fairly flat spectrum that is not consistent 
with a T dwarf or an L dwarf.
\end{landscape}
}

\section{Multi-wavelength photometric follow-up}
\label{VHS_dT:phot_obs}

In this section, we describe the optical and near-infrared photometric 
follow-up carried out prior to spectroscopic observations to confirm the 
substellar status of our candidates (except for VHS J023756.24$-$063142.9 
which is covered by SDSS DR8).

\subsection{Optical photometry}
\label{VHS_dT:phot_obs_Optical}

We carried out optical photometry of two of the brightest T dwarf candidates 
with the Johnson $I$, and Sloan $i,z$ filters available on CAMELOT 
installed on the 0.82-m IAC80 telescope. CAMELOT (CAmara MEjorada del 
Observatorio del Teide or Improved Camera in the Teide Observatory in 
English) is a wide-field camera equipped with a 2048$\times$2048 pixels 
Charge-Coupled Device (CCD) with 13.5 micron pixels, resulting in a pixel 
scale of 0.304 arcsec and a field-of-view of 10.6 arcmin squared. 

We obtained two images of 20 min on-source integrations in the Sloan $i$ 
filter on 3 May 2011 for VHS J162208.94$-$095934.6\@. Additionally, we
observed this object in Johnson $I$ (3$\times$15 min) and in Sloan $z$
(3$\times$15 min) on 5 May 2011 along with a photometric standard star 
\citep[G153-41;][]{landolt09} to calibrate our Johnson $I$ image. The seeing
was around 1.5--1.8 arcsec. Note that neither our objects nor G153-41 are
covered by Sloan so we do not have any photometric zero points for the
Sloan $z$ filter on that specific night. The data reduction of the optical
images taken with CAMELOT was standard and made automatically by a pipeline
at the end of each observing night. It includes bias subtraction and 
flat-field correction. No astrometry was performed because we are primarily
interested in the photometry (detection or lower limit).

We did not detect any object in the combined $i$-band image of
VHS J162208.95$-$095934.6 down to a 3$\sigma$ detection limit of 
21.48$\pm$0.12 mag (AB system), implying an $i-J$ colour redder than 
5.26$\pm$0.12 mag, consistent with a spectral type later than L6--L7
\citep{schmidt10b}. This detection limit is taken from the rms of the number 
counts at the expected position of the object and compared to nearby objects
with known magnitudes calibrated with a Sloan standard field 
(RA\,=\,17:00:29, dec\,=\,$+$33:58:51; 3$\times$30 sec; airmass=1.16)
observed in Sloan $i$ on the same night. The dispersion of the photometric
calibration using a large number of stars in the standard field is 0.12 mag. 
We detected the target marginally in the Johnson $I$ and infer a 
magnitude of 20.86$\pm$0.50 mag (Vega system), implying an $I-J$ colour 
of $\sim$4.64, consistent with T dwarfs \citep{liebert06}. We also marginally
detected the object in SDSS$z$ with $Z$\,=\,20.72$\pm$0.11 mag (2$\sigma$ 
detection) using the zero point from the night of 7 May 2011 (see below), 
confirming that it is an interesting cool brown dwarf candidate. 

Moreover, we obtained four exposures of 15 min in Johnson $I$ and six images
of 10 min in Sloan $z$ on 7 May 2011 with CAMELOT with a seeing of 1.7--2.0
arcsec for VHS J154352.78$-$043909.6\@. We derive magnitudes of 
$I$\,=\,22.02$\pm$0.15 mag (Johnson filter; Vega system) and 
$z$\,=\,20.95$\pm$0.27 mag (AB system; 2$\sigma$ confidence limits). 
The Sloan $z$ zero points come from the observations of two photometric 
standards \citep[PG\,0918$+$029 and PG\,1528$+$062][]{landolt09} on the same 
night with a calibration that agrees to better
than 0.1 mag. Hence, the $I-J$ and $z-J$ colours suggest that this candidate 
is a good mid- to late-T dwarf suitable for spectroscopic follow-up.

In addition to the IAC80 optical follow-up, we targeted the most interesting 
of our targets with the Sloan $z$ filter available on the Laica instrument 
mounted on the prime focus of the Calar Alto 3.5-m telescope (Director's
Discretionary Time). This target was too faint for the IAC80 telescope.
Laica uses a 2$\times$2 mosaic of 4096$\times$4096 CCDs pixels offering a 
field-of-view of 44.36 arcmin square with a 0.224 arcsec pixel scale. 
The night of 12 May 2011 was not photometric because of clouds passing by,
but the seeing was below 1 arcsec during the observations. We took 18 images 
of 100 sec with a manual dither pattern, yielding a combined image of 30 min. 
The known T7 dwarf, SDSS J150411.63$+$102718.4 \citep{chiu06}, was observed 
with the same set-up to photometrically calibrate the Laica images with all 
stars present in that Sloan calibration field. The data reduction was 
standard for optical imaging. The target is not detected down to a 3$\sigma$ 
detection limit of $z$\,=\,21.9 mag (AB system), implying a $z-J$ colour 
redder than 2.76 mag, which classified this object as a potential T dwarf 
\citep{hawley02,pinfield08,schmidt10b}.

\subsection{Near-infrared photometry}
\label{VHS_dT:phot_obs_NIR}

We conducted near-infrared photometry with Omega2000 on the Calar Alto
3.5-m telescope to detect the most interesting-looking candidate in the
$H$-band (VHS 5$\sigma$ limit is 18.71$\pm$0.40 mag) and confirm its T dwarf 
status through methane imaging.

Omega2000 employs a HAWAII-2 2048$\times$2048 pixel detector sensitive
in the 1--2.5 micron wavelength range with a pixel size of 0.45 arcsec, 
yielding a field-of-view of 15.4 by 15.4 arcmins \citep{kovacs04}. 

We imaged VHS J143311.46$-$083736.3 on 22 June 2011 with Omega2000 for 90 min 
in methane off (central wavelength is 1.579 microns with a 
full-width-at-half-maximum of 0.101 microns) and 30 min in $H$ on 23 June 2011 
(between 20h30 till 21h10 UT at an airmass of $\sim$1.12) along with a SDSS 
T7 dwarf \citep[SDSS J150411.63$+$102718.4 (T7); $H$\,=\,16.909;][]{chiu06}
as reference for photometric calibration at an airmass of 1.45\@.
The total exposure time was divided into five-point dither patterns with
individual on-source integrations of 6$\times$10s and 3$\times$20s in $H$ 
and methane off (CH$_{4 \rm off}$), respectively, for both sources. The 
presence of clouds during the first night hampered the methane observations 
of our T dwarf candidate, resulting in 50 min worth of data instead of the 
90 min on-target. The second night was also not photometric, with variations
in the level of the sky of up to 40\%. The seeing on the combined $H$-band 
and methane images of 
VHS J143311.46$-$083736.3 was 0.6--0.9 and 1--1.3 arcsec, respectively.
The total exposure times in $H$ and methane for SDSS J150411.63$+$102718.4
were set to 5 and 20 min, respectively.

We reduced the near-infrared data in a standard manner under the 
IRAF\footnote{IRAF 
is distributed by the National Optical Astronomy Observatory, which is 
operated by the Association of Universities for Research in Astronomy (AURA) 
under cooperative agreement with the National Science Foundation} environment.
Series of five dithered positions were observed in each filter: for
each position in each filter, we subtracted a median of the other four frames
used to create a sky frame and then divided by the normalised combined dome 
flat (ON\,$-$\,OFF exposures). Then, we averaged the five frames of each cube 
with a ccdclip rejection algorithm and again averaged the various cubes 
to create a final combined image for both sources in both filters.

We calibrated astrometrically the Omega2000 images with the astrometry.net
package\footnote{More details at astrometry.net}, which requires the centre
of the image given by the (RA,dec) coordinates in sexagesimal format, the 
pixel scale (0.45 arcsec/pixel with an allowance of 0.03 arcsec/pixel), and 
a radius for the search (set to 30 arcmin, i.e.\ twice the field-of-view 
of Omega2000 images). We found that the astrometric solution was satisfactory 
and of good quality for our purposes, comparing with 2MASS 
\citep{cutri03,skrutskie06}, the UKIDSS Large Area Survey catalogue 
\citep{lawrence07} for SDSS J150411.63$+$102718.4, and the VHS database for 
VHS J143311.46$-$083736.3\@.

We extracted the photometry of our targets with the GAIA tool\footnote{GAIA 
--- the Graphical Astronomy and Image Analysis Tool --- is a derivative of 
the Skycat catalogue and image display tool, developed as part of the VLT 
project at ESO. Skycat and GAIA are free software under the terms of the 
GNU copyright. See http://star-www.dur.ac.uk/$\sim$pdraper/gaia/
for more details}, which itself uses SExtractor \citep{bertin96}.
We ran the detection algorithm to extract all sources whose flux was
at least 2$\sigma$ above the threshold and detection background.
We calibrated photometrically the images of the template T dwarf and
our target with the UKIDSS and VHS images, respectively, because they
have more sources in common and higher photometric accuracy 
than 2MASS\@. We found a difference of 0.14 mag between the zero points
in the $H$-band for both objects. We derived a revised magnitude of
$H$\,=\,16.768$\pm$0.024 mag for SDSS J150411.63$+$102718.4, compared 
to the upper limit of $H$\,=\,16.91 mag from 2MASS \citep{chiu06}. We detected
VHS J143311.46$-$083736.3 on the Omega2000 combined images and 
inferred $H$\,=\,19.22$\pm$0.14 mag. We calibrated photometrically
the methane photometry by comparing with the $H$ photometry and assuming
that the $H$--CH$_{4 \rm off}$ colour should be neutral. We inferred colours
of 0.45$\pm$0.03 mag (5$\sigma$ detection) and 0.67$\pm$0.20 
(1.8$\sigma$ confidence) for SDSS J150411.63$+$102718.4 and 
VHS J143311.46$-$083736.3, respectively. We emphasise that
our T dwarf candidate is one of the faintest objects on the image,
resulting in a poor detection in both filters. Nonetheless, the high
value for the $H$--CH$_{4 \rm off}$ colour of our candidate suggests
a spectral type of T7 or later \citep{goldman10}.

\section{Spectroscopic follow-up}
\label{VHS_dT:spec_obs}

In this section, we describe the near-infrared spectroscopic follow-up 
conducted with VLT/X-shooter and Magellan/FIRE\@.

\subsection{VLT X-shooter}
\label{VHS_dT:spec_obs_X-shooter}

We obtained spectroscopic observations for two candidate VHS T dwarfs with 
X-shooter on the Very Large Telescope on 5 and 8 June 2011, under the ESO 
programme 087.C-0639(A). We used the echelle slit mode, which covers the 
wavelength range 300--2500nm. This is split into three separate arms, 
the UVB (ultraviolet; 300--550nm), VIS (visible; 550--1000nm) and NIR 
(near-infrared; 1000--2500nm). Using slit 
widths of 1.0 arcsec for the UVB arm and 0.9 arcsec for the VIS and NIR arms,
the resolution of the VIS and NIR arms are R\,=\,8800 and R\,=\,5600, 
respectively. We took four individual integrations in an ABBA pattern, 
totalling to integration times of 920s in the UVB arm, 1200s in the VIS arm 
and 1560s in the NIR arm for VHS J162208.9$-$095934.6\footnote{This object 
was independently published by \citet{kirkpatrick11} and classified as T6\@.} 
and 1720s in the UVB arm, 2000s in the 
VIS arm and 2360s in the NIR arm for VHS J92934.1$-$442550.5\@. We note 
that we do not expect any significant flux in the UVB wavelength range for 
members of our sample and therefore we do not show the details for this arm. 
We observed telluric standard stars before or after every target at similar 
airmass. Sky flats and arc frames were also taken at the beginning of every 
night.

\begin{figure}
  \centering
  \includegraphics[width=\linewidth, angle=0]{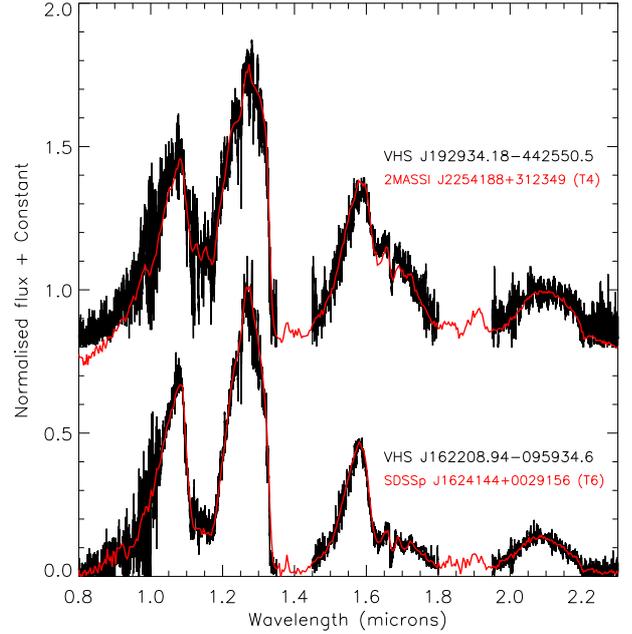}
  \caption{
Low-resolution near-infrared spectra obtained with VLT X-shooter and 
normalised at 1.265 microns. 
Overplotted in red are the known T-dwarf templates that best fit our spectra: 
2MASSI J2254188$+$312349 \citep[T4;][]{burgasser04c} and
SDSSp J162414.37+002915.6 \citep[T6;][]{burgasser06b}.
}
  \label{fig_VHS_dT:spectra_dT_XSH}
\end{figure}

The data were reduced using the ESO X-shooter pipeline (version 1.3.7). 
The pipeline removes non-linear pixels, subtracts the bias (in the VIS arm) or 
dark frames (in the NIR arm) and divides the raw frames by flat fields. Images 
are pair-wise subtracted to remove sky background. The pipeline then extracts 
and merges the different orders in each arm, rectifying them using a 
multi-pinhole arc lamp (taken during the day-time calibration) and 
corrects for the flexure of the instrument using single-pinhole arc lamps 
(taken at night, one for each object observed). Telluric stars are reduced 
in the same way, except that sky subtraction was performed by fitting the 
background (as tellurics are not observed in nodding mode). The spectra 
were telluric-corrected and flux-calibrated using IDL routines, following 
a standard 
procedure: first the telluric spectrum is cleared of HI absorption lines 
(by interpolating over them) and scaled to match the measured magnitudes; 
then it is divided by a blackbody curve for the appropriate temperature to 
obtain the instrument$+$atmosphere response curve; finally the target 
spectra is multiplied by the response curve obtained to flux-calibrate it.
The arms (VIS and NIR) were then merged by matching the flux level in the
overlapping regions between them. Finally, each spectrum was visually 
inspected to check for possible problems during the extraction or merging 
stage. The final reduced spectra are displayed in black in
Fig.\ \ref{fig_VHS_dT:spectra_dT_XSH} along with spectral templates
to assign spectral types visually \citep{burgasser04c,burgasser06b}.

\begin{figure}
  \centering
  \includegraphics[width=\linewidth, angle=0]{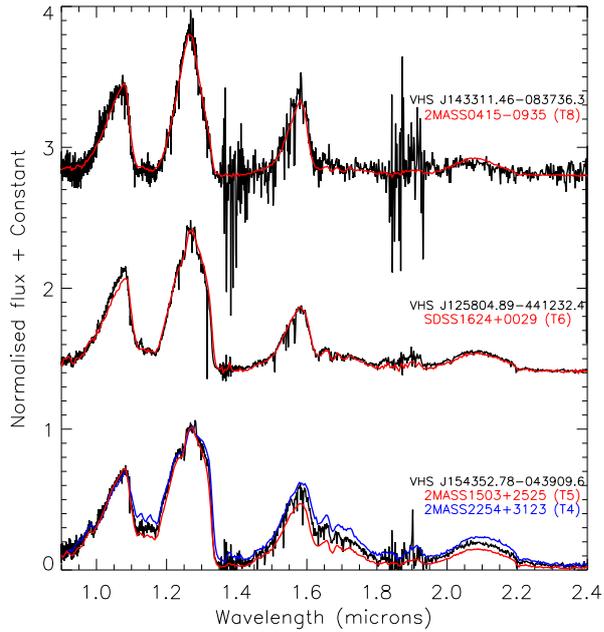}
  \caption{
Low-resolution near-infrared spectra obtained with Magellan/FIRE and 
normalised at 1.265 microns. 
Overplotted in red are the known T-dwarf templates that best fit our spectra:
2MASS J15031961+2525196 \citep[T5;][]{burgasser04c},
SDSSp J162414.37+002915.6 \citep[T6;][]{burgasser06b}, and
2MASSI J0415195-093506 \citep[T8;][]{burgasser04c}, and in blue
the T4 dwarfs, 2MASSI J2254188$+$312349 \citep{burgasser04c}.
}
  \label{fig_VHS_dT:spectra_dT_FIRE}
\end{figure}
\subsection{Magellan/FIRE}
\label{VHS_dT:spec_obs_FIRE}

The Folded port InfraRed Echellette (FIRE) spectrograph \citep{simcoe08,simcoe10} 
mounted on the Baade 6.5m Magellan telescope and Las Campanas Observatory 
was used to obtained near-infrared spectroscopy for four VHS T dwarf candidates. 
VHS J20515938$-$5508439 was observed on 17 August 2011 with a 1 arcsec 
slit in prism mode. VHS J12580489$-$4412324, 
VHS J14331146$-$0837363, and VHS J15435278$-$0439096 were all observed on the 
6 March 2012, using a 0.8 arcsec slit, also in prism mode. Both set-ups 
deliver similar R\,$\sim$\,200--300\@. All science observations 
were obtained using the sample up the ramp (SUTR) detector read-out mode. 
Telluric standard stars of type A0V were observed at similar airmass to the 
science targets and arc frames were taken in situ.

The data were reduced using the low-dispersion version of the FIREHOSE pipeline, 
which is based on the MASE pipeline developed for the Magellan Echellette
optical spectrograph \citep{bochanski09,simcoe10}. The pipeline 
uses a flat field constructed from two quartz lamp images taken with the lamp 
at high (2.5 V) and low (1.5 V) voltage settings. The data were divided by 
this pixel flat before they were wavelength calibrated, and the object spectra 
identified. The pipeline performs sky subtraction following the method outlined 
in \citet{bochanski11}, adapted for the low-dispersion configuration of the 
spectrograph. The spectra were optimally extracted before being combined using 
an adaptation of the {\tt{xcombspec}} routine from SpexTool \citep{cushing04}. 
The T dwarf spectra were then corrected for telluric absorption and 
flux-calibrated using a FIRE specific version of the {\tt{xtellcor}} routine 
\citep{vacca03}. Finally, residual outlying points caused by cosmic rays were
removed using a simple sigma-clipping algorithm.

The final reduced spectra are displayed in black in 
Fig.\ \ref{fig_VHS_dT:spectra_dT_FIRE} where they are compared to T dwarf
spectral templates shown in red \citep{burgasser04c,burgasser06b}.

\section{Properties of the new T dwarfs}
\label{VHS_dT:new_dT}

In this section, we assign spectral types to the new T dwarfs and
derive their spectroscopic distances.

\longtab{2}{
\begin{landscape}
 \begin{longtable}{@{\hspace{0mm}}c @{\hspace{2mm}}c @{\hspace{2mm}}c @{\hspace{2mm}}c @{\hspace{2mm}}c @{\hspace{2mm}}c @{\hspace{2mm}}c @{\hspace{2mm}}c @{\hspace{2mm}}c@{\hspace{0mm}}}
 \caption{\label{tab_VHS_dT:tab_ratios} Spectral ratios (unitless) measured for our confirmed T dwarfs} \\
 \hline
 \hline
VHS J\ldots{}  & H20J (SpT) & CH4J (SpT) & Wj (SpT) & H2OH (SpT) & CH4H (SpT) &  NH3H (SpT) & CH4K (SpT) & K/J \cr
 \hline
  \endfirsthead
  \caption{continued.} \\
  \hline
  \hline
VHS J\ldots{}  & H20J (SpT) & CH4J (SpT) & Wj (SpT) & H2OH (SpT) & CH4H (SpT) &  NH3H (SpT) & CH4K (SpT) & K/J \cr
  \hline
  \endhead
  \hline
  \endfoot
125804.89$-$441232.4 & 0.155$\pm$0.001 (T6)     & 0.277$\pm$0.001 (T7) & 0.451$\pm$0.001 ($<$T7) & 0.295$\pm$0.002 (T6)    & 0.352$\pm$0.002 (T6) & 0.694$\pm$0.003 ($<$T7.5) & 0.173$\pm$0.004 (T6)     & 0.154$\pm$0.001 \cr
143311.46$-$083736.3 & 0.027$\pm$0.006 ($>$T7)  & 0.147$\pm$0.003 ($>$T7) & 0.299$\pm$0.003 (T8)    & 0.060$\pm$0.010 ($>$T9) & 0.179$\pm$0.007 (T7) & 0.562$\pm$0.010 (T8--T9)  & 0.131$\pm$0.038 (T6--T7) & 0.087$\pm$0.003  \cr
154352.78$-$043909.6 & 0.294$\pm$0.003 (T5)     & 0.430$\pm$0.002 (T5) & 0.608$\pm$0.002 ($<$T7) & 0.364$\pm$0.004 (T5)    & 0.518$\pm$0.003 (T4) & 0.780$\pm$0.005 ($<$T7.5) & 0.274$\pm$0.006 (T4)     & 0.195$\pm$0.001  \cr
162208.94$-$095934.6 & 0.190$\pm$0.089 (T5--T6) & 0.333$\pm$0.055 (T6) & 0.488$\pm$0.151 ($<$T7) & 0.304$\pm$0.032 (T6)    & 0.292$\pm$0.030 (T6) & 0.705$\pm$0.066 ($<$T7.5) & 0.179$\pm$0.213 (T5--T6) & 0.152$\pm$0.017  \cr
192934.18$-$442550.5 & 0.342$\pm$0.354 (T3--T5) & 0.464$\pm$0.770 (T4) & 0.640$\pm$1.164 ($<$T7) & 0.394$\pm$0.021 (T4)    & 0.631$\pm$0.031 (T3) & 0.750$\pm$0.067 ($<$T7.5) & 0.398$\pm$0.061 (T3)     & 0.212$\pm$0.144  \cr
 \hline
\end{longtable}
\end{landscape}
}

\subsection{Spectral types}
\label{VHS_dT:spec_SpT}

We derived spectral types for our new T dwarfs using templates available in 
the literature. We used the database of T dwarf templates\footnote{Data
available at http://pono.ucsd.edu/$\sim$adam/browndwarfs/spexprism/} 
from the 0.8-5.5 Micron Medium-Resolution Spectrograph and Imager (SpeX) 
on the 3-m NASA Infrared Telescope Facility (IRTF) to estimate spectral
types for our targets with an uncertainty of half a subclass. These templates
were selected from a sample of 43 T dwarfs (known at that time) as part 
of the unified near-infrared classification scheme designed by 
\citet{burgasser06a}. We compared the VLT/X-shooter and Magellan/FIRE spectra 
of our objects to the following T dwarfs observed with IRTF/SpeX 
(Figs \ref{fig_VHS_dT:spectra_dT_FIRE} and \ref{fig_VHS_dT:spectra_dT_XSH}): 
2MASSI J2254188$+$312349, 2MASS J15031961+2525196, 2MASSI J0415195-093506 
\citep[T4, T5, and T8;][]{burgasser04c} and SDSSp J162414.37+002915.6 
\citep[T6;][]{burgasser06b}. Adopted spectral types, accurate to half 
a subclass, are listed in Table \ref{tab_VHS_dT:tab_new_dT}.

Finally, we note that we adopted a spectral type of T6.0$\pm$0.5 for
VHS J162208.9$-$095934.6, in agreement with \citet{kirkpatrick11}.
All new VHS T dwarfs agree well with the SpeX spectral templates. The 
only exception may be VHS J125804.89$-$441232.4, whose flux in the $Y$-band
is above the T6 standard \citep{strauss99,tinney03,burgasser06a}.
The $Y$-band is probing lower layers of the atmospheres so an increased
flux at these wavelengths may indicate the presence of less dust or decreased 
opacity. A lower metallicity could also be responsible for the increased flux 
in the $Y$-band, in agreement with the low-metallicity trends seen in other 
T dwarfs \citep{burgasser10b,kirkpatrick11}.

Moreover, we present near-infrared spectral typing flux ratios for five
of our T dwarfs (Table \ref{tab_VHS_dT:tab_ratios}), following definitions
described in \citet{burgasser06a}, \citet{warren07c}, and \citet{delorme08a}.
The spectral types derived from those spectral indices agree well with our
spectral classification from the direct comparison with T dwarf templates.
We note that no spectral types are derived from the K/J ratio because
\citet{burgasser06a} noted a high dispersion in this ratio, suggesting
that gravity and metallicity should be incoporated in a future classification
scheme \citep{kirkpatrick08,cruz09}. Furthermore, the Wj and NH3H ratios
defined by \citet{warren07c} and \citet{delorme08a}, resepctively, are mainly
designed for T dwarfs later than T7.5--T8, placing little constraints on
our five new T dwarfs with earlier spectral types but working fairly
well for VHS J143311.46$-$083736.3\@.

\subsection{Spectroscopic distances}
\label{VHS_dT:spec_dist}

Several spectral type vs absolute magnitude relations have been proposed
over the years and improved progressively with the larger number of
L and T dwarfs with parallaxes. In this paper, we use the latest relations
between $JHK$ magnitudes and spectral types in the L0--T8 range based on
trigonometric parallaxes as
derived by \citet{faherty12} in their Table 7\@. The dispersion for
these relations range from 0.27 ($K$-band) to 0.3 ($J$-band).
Earlier relations can be found in \citet{dahn02}, \citet{tinney03}
\citet{vrba04}, \citet{liu06}, \citet{marocco10}, and \citet{kirkpatrick12}.
We should point out that the parallax of 2MASSI J0415195$-$093506 
\citep{vrba04} is best reproduced by the relations in \citet{faherty12},
with the \citet{marocco10} relations producing a result that is 6.3\% too low.

With these relations, we derive absolute magnitudes of
$J$\,=\,14.632$\pm$0.093, 14.725$\pm$0.121, 15.199$\pm$0.243,
15.738$\pm$0.355, and 16.515$\pm$0.422 mag for T4, T4.5, T6, T7, and T8
dwarfs, respectively. The uncertainty is the upper limit on the error, assuming 
an accuracy of 0.5 spectral type. We estimated the spectroscopic distance
of 2MASSI J0415195$-$093506 \citep{burgasser02} to be 5.83$\pm$1.26 pc, 
in agreement within the error bars with its trigonometric distance
\citep[5.736$\pm$0.092 pc;][]{vrba04}. We derived distances within 25 pc for
three of our T dwarfs, VHS J125804.9$-$441232.4 (T6), VHS J162208.9$-$095934.6
(T6), and VHS J154352.8$-$043909.6 (T4.5), hence adding three new T dwarfs
in the solar vicinity. The remaining T dwarfs are farther away in the
30--50 pc range. VHS J205159.4$-$550843.9 is the faintest of all T7
dwarfs published to date in the $J$-band with $J$\,=\,19.280$\pm$0.083 mag,
suggesting that it is the most distant of its spectral class. The record 
holder was ULAS J134940.81$+$091833.3 with $J$\,=\,19.16 mag
\citep{burningham10b} although another T7 dwarf was classified 
photometrically (i.e.\ no spectrum available yet) as a T7 and is two 
magnitudes fainter with $J$\,=\,20.143$\pm$0.040 mag \citep{lodieu09b}.

All numbers and results discussed above are summarised in
Table \ref{tab_VHS_dT:tab_new_dT}. We assumed that the new T dwarfs
are single, keeping in mind that the binary fraction of late-L/early-T 
(known as transition objects) and T dwarfs is about
31$^{+21}_{-15}$\% and 14$^{+14}_{-7}$\%, respectively 
\citep{burgasser07a,looper08a,goldman08}. 
Moreover, we assumed that the near-infrared extinction is negligible 
along the line of sight of our T dwarfs.

\subsection{Proper motion measurements}
\label{VHS_dT:spec_PM}

Proper motion measurements for our candidates turned out to be difficult 
because the epoch difference between the VHS DR1 and WISE data is only
about a few months. However, for three brighter candidates 
(Table \ref{tab_VHS_dT:tab_new_dT}) we succeded in measuring their proper 
motions after searching via the 
CDS\footnote{http://vizier.u-strasbg.fr/cgi-bin/VizieR} and finding faint 
counterparts in 2MASS \citep{skrutskie06} observed about 11 years before 
VHS and WISE\@.
Two of them (VHS J125804.9$-$441232.4 and VHS J154352.8$-$043909.6) have 
only $J$-band detections in 2MASS, whereas VHS J162208.9$-$095934.6 was 
detected in both the 2MASS $J$- and $K_s$-bands.
In addition, we identified one of them VHS J125804.9$-$441232.4 with a 
$J$-only detection in DENIS \citep{epchtein97}, about three years earlier 
than 2MASS\@.

For our proper motion solutions listed in Tab.~\ref{tab_VHS_dT:tab_new_dT} we 
applied a weighted linear fitting over all available multi-epoch positions 
($N_{\rm ep}$), a method similar to that of \citet{scholz12a}. We assumed 
70 mas errors 
for the individual VHS DR1 positions taken from the catalogues in the VISTA
Science Archive and 150 mas errors for the faint object 
detections in 2MASS and DENIS\@. For the VHS we expect similar astrometric
errors to those of UKIDSS, varying between 50 and 100 mas depending 
on Galactic latitude \citep{lawrence07}. For 2MASS and DENIS, we assumed
those astrometric errors based on former estimates by \citet{scholz12a}, 
\citet{liu11b}, and \citet{kirkpatrick11}.
For the WISE all-sky release positions of VHS J125804.9$-$441232.4 and 
VHS J154352.8$-$043909.6 we used the positional errors given by WISE (about 
100 and 350 mas, respectively)\footnote{We favoured the WISE all-sky release
for the proper motion determination.}. In case of VHS J162208.9$-$095934.6 we 
preferred to use the two-epoch positions from WISE with their errors (150--170 mas)
as given in \citet{kirkpatrick11}, who also provided two additional epoch 
positions from \textit{Spitzer} (with errors of 130--210 mas) that we also 
included in our proper motion fitting. Our proper motions for 
VHS J162208.9$-$095934.6 are consistent with those of 
\citet{kirkpatrick11}, but about twice as precise.

We also tried to obtain proper motions for the other candidates, including
VHS J023756.2$-$063142.9 with its assumed SDSS DR8 counterpart and
VHS J1433114$-$083736.3 with its additional Omega2000 position, but failed 
to determine consistent results. Therefore, we assume that the SDSS 
counterpart of VHS J023756.2$-$063142.9 may be wrong, whereas the proper 
motion of VHS J1433114$-$083736.3 is probably shorter than about 50 mas/yr.

\section{Conclusions}
\label{VHS_dT:conclusions}

We have presented the near-infrared spectra of six new T dwarfs identified in
675 square degrees common to the southern sky imaged by the VISTA Hemisphere 
Survey and the Wide Field Infrared Survey Explorer satellite. One of them
was independently reported as a T6 dwarf while another one was known 
previously. The other sources remain photometric T dwarf candidates because
we could not obtain near-infrared spectroscopy yet. Three confirmed
T dwarfs lie within the solar vicinity, i.e.\ within the 25 pc limit.
The only T7 dwarf in our sample has the faintest $J$-band magnitude
of all T7 dwarfs published to date. We also improved the proper motion
measurements of the brightest T dwarfs with data from various public surveys.

We have shown the potential of the VHS in complementing mid-infrared 
photometry from the WISE mission, but we stress that an all-sky
optical ($i$ and/or $z$) survey of the entire southern sky would optimize
our searches significantly, as is the case for UKIDSS with the Sloan
coverage.  We plan to expand our searches to the WISE 
all-sky release and the ever increasing area of the sky imaged by VHS up to 
its completion to (1) discover cool brown dwarfs, in particular the Y dwarfs 
recently announced by \citet{cushing11} and \cite{kirkpatrick12}, 
(2) provide a magnitude-limited  census of T dwarfs over the entire southern 
sky at completion of the VHS public survey, (3) discover gravity/metallicity 
benchmarks \citep[e.g.][]{pinfield12}, (4) constrain the field mass function 
at low temperatures with high accuracy \citep{kroupa02,chabrier03} as 
attempted by several groups with large-scale surveys 
\citep{metchev08,burningham10b,reyle10,kirkpatrick12}.

\begin{acknowledgements}
NL was funded by the Ram\'on y Cajal fellowship number 08-303-01-02 and the 
national program AYA2010-19136 funded by the Spanish ministry of science and 
innovation. ADJ is supported by a FONDECYT Postdoctorado under project number
3100098\@. ADJ is also partially supported by the Joint Committee 
ESO-Government of Chile. JJ is also supported by a FONDECYT postdoctorado 
fellowship (project number 3110004). ADJ and MTR acknowledge the 
support of the grant from CONICYT and the partial support from Center for 
Astrophysics FONDAP and Proyecto Basal PB06 (CATA).
This research has been supported by the Spanish Ministry of Economics and 
Competitiveness and the ``Fondo Europeo de Desarrollo Regional'' FEDER under
the project AYA2010-21308-C03-02, AYA2010-21308-C03-03, and
AYA2010-20535\@. PC is funded by RoPACS\@. NL, ADJ, and DP have received
funding from RoPACS\@.

Based on observation obtained as part of the VISTA Hemisphere Survey, 
ESO Progam, 179.A-2010 (PI: McMahon).
The VISTA Data Flow System pipeline processing and science archive are 
described in \cite{irwin04} and \citet{hambly08}. We have used data from the 
first data release, which is described in detail in McMahon et al.\ (2012).
We are grateful to the Cambridge Astronomy Survey Unit and the VISTA Science 
Archive at the Wide Field Astronomy Unit, Edinburgh, for providing us with 
the reduced data and catalogues.

This article is based on observations made with the 0.82-m IAC80 telescope
operated on the island of Tenerife by the IAC in the Spanish Observatorio 
del Teide. Based on data obtained with the Omega2000 wide-field camera 
installed on the Calar Alto 3.5-m telescope. 
The X-Shooter spectroscopy is based on observations made with ESO telescopes 
at the La Silla Paranal Observatory under programme ID 089.C-0854 in visitor
mode. This paper includes data gathered with the 6.5 meter Magellan Telescopes 
located at Las Campanas Observatory, Chile.

This research has made use of the Simbad and Vizier databases, operated
at the Centre de Donn\'ees Astronomiques de Strasbourg (CDS), and
of NASA's Astrophysics Data System Bibliographic Services (ADS).

This publication makes use of data products from the Two Micron
All Sky Survey (2MASS), which is a joint project of the University
of Massachusetts and the Infrared Processing and Analysis
Center/California Institute of Technology, funded by the National
Aeronautics and Space Administration and the National Science Foundation.

Funding for the SDSS and SDSS-II has been provided by the Alfred P. Sloan
Foundation, the Participating Institutions, the National Science Foundation,
the U.S. Department of Energy, the National Aeronautics and Space 
Administration, the Japanese Monbukagakusho, the Max Planck Society, and the 
Higher Education Funding Council for England. The SDSS Web Site is 
http://www.sdss.org/. The SDSS is managed by the Astrophysical Research 
Consortium for the Participating Institutions. The Participating Institutions 
are the American Museum of Natural History, Astrophysical Institute Potsdam, 
University of Basel, University of Cambridge, Case Western Reserve University, 
University of Chicago,
Drexel University, Fermilab, the Institute for Advanced Study, the Japan
Participation Group, Johns Hopkins University, the Joint Institute for Nuclear
Astrophysics, the Kavli Institute for Particle Astrophysics and Cosmology, the
Korean Scientist Group, the Chinese Academy of Sciences (LAMOST), Los Alamos
National Laboratory, the Max-Planck-Institute for Astronomy (MPIA), the
Max-Planck-Institute for Astrophysics (MPA), New Mexico State University, Ohio
State University, University of Pittsburgh, University of Portsmouth, Princeton
University, the United States Naval Observatory, and the University of Washington.

This publication makes use of data products from the Wide-field Infrared
Survey Explorer, which is a joint project of the University of California,
Los Angeles, and the Jet Propulsion Laboratory/California Institute of
Technology, funded by the National Aeronautics and Space Administration.
\end{acknowledgements}

  \bibliographystyle{aa}
  \bibliography{../mnemonic,../biblio_old}

\end{document}